\documentclass[%
 aip,
 amsmath,amssymb,
 reprint,%
floatfix
]{revtex4-1}

\usepackage{graphicx}
\usepackage{dcolumn}
\usepackage{bm}

\usepackage[utf8]{inputenc}
\usepackage[T1]{fontenc}
\usepackage{mathptmx}

\begin{document}

\preprint{AIP/123-QED}

\title{Electron Beam Irradiation of Gallium Nitride-on-Silicon Betavoltaics Fabricated with a Triple Mesa Etch}

\author{T. Heuser}
 \altaffiliation{Corresponding author, email: theuser@stanford.edu.}
 
\author{M. Braun}%

\affiliation{ 
Department of Materials Science and Engineering, Stanford University, 450 Serra Mall, Stanford, CA 94305
}%

\author{P. McIntyre}%

\affiliation{ 
Department of Materials Science and Engineering, Stanford University, 450 Serra Mall, Stanford, CA 94305
}

\author{D.G. Senesky}
\affiliation{%
Department of Aeronautics and Astronautics, Stanford University, 450 Serra Mall, Stanford, CA 94305
}%
\affiliation{Department of Electrical Engineering, Stanford University, 450 Serra Mall, Stanford, CA 94305}
 
\date{\today}

\begin{abstract}
A process for growing gallium nitride (GaN) vertical p-i-n homojunctions on (111) silicon (Si) substrates using metalorganic chemical vapor deposition (MOCVD) was developed, and a triple mesa etch technique was used to fabricate efficient betavoltaic energy converters.  Monte Carlo simulation platform CASINO was used to model beta radiation penetration into GaN to aid device design. The resulting devices were tested under irradiation from a scanning electron microscope (SEM) electron beam (e-beam) tuned to imitate the energies of the \textsuperscript{63}Ni beta emission spectrum. Based on current-voltage (I-V) measurements taken under e-beam illumination, a maximum open-circuit voltage of 412 mV and a maximum short-circuit current density of 407 nA/cm\textsuperscript{2} were measured. A high fill factor (FF) of 0.77 and power conversion efficiency of 6.6\% were obtained. Additionally, the proposed triple mesa etch technique used to create these betavoltaics has potential for further use in fabricating many types of electronic devices using a wide variety of material platforms.

\end{abstract}

\maketitle

\section{Introduction}

Betavoltaic energy converters are semiconductor devices that harvest energy from decaying radioisotopes.  Unlike radioisotope thermoelectric generators (RTGs), which harvest heat produced by decaying radioisotopes, betavoltaics convert the energy from ionizing radiation directly into electricity. Originally developed in the mid-1950s, early betavoltaics were silicon (Si) devices powered by \textsuperscript{147}Pm, and were used to power implantable pacemakers. \cite{paul1956radioactive,olsen1972nuclear}

Betavoltaics have significant potential as an extremely versatile power source for electronic devices in situations requiring small size, long operational lifetimes, in which conventional means of power generation are difficult, or in locations where harsh environments (high temperature, radiation-rich, corrosive, etc.) preclude the use of other types of power sources.  Examples include remote sensor platforms for space, deep sea, deep underground, biomedical implants, and in-situ structural health monitoring.\cite{honsberg2005gan, park2008energy}

\begin{figure}[t]
\centerline{\includegraphics[width=\columnwidth]{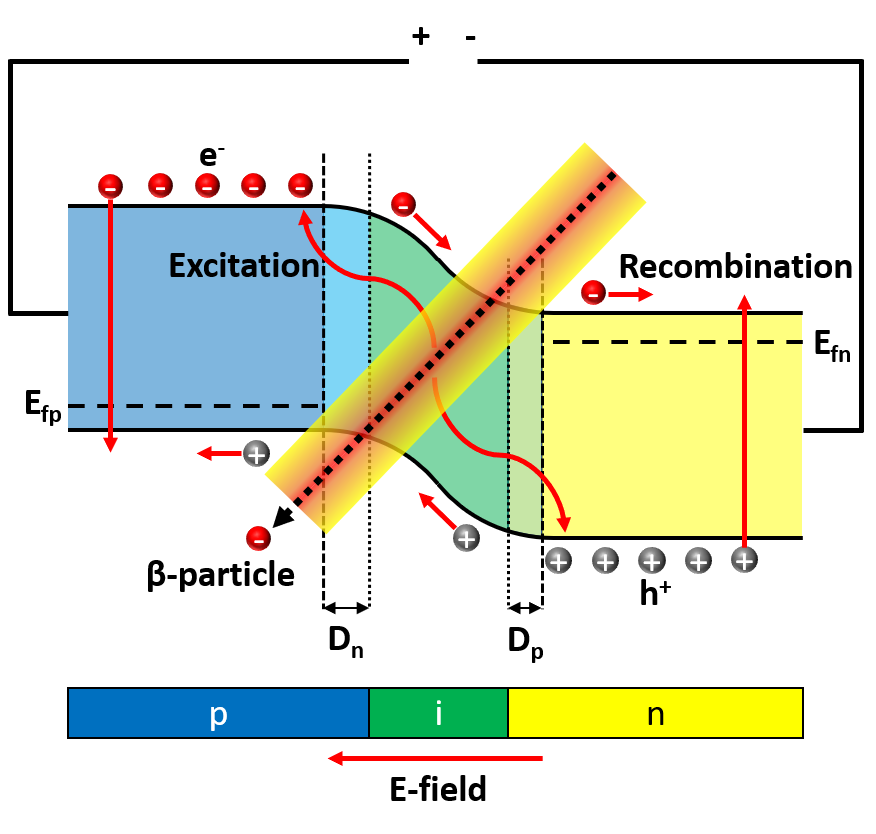}}
\caption{Band diagram showing charge carrier excitation and recombination as the result of a beta particle strike. Electrons and holes produced in the intrinsic region or one minority carrier diffusion length on either side (represented by D\textsubscript{n} and D\textsubscript{p}) contribute to the generated current. Region widths are not to scale.}
\label{band diagram}
\end{figure}

The physics principles behind the operation of betavoltaics are very similar to those of photovoltaics.  Both utilize the electric field built up across the depletion region of a semiconductor junction to capture electrons and holes generated by incident energetic particles. However, photovoltaics are powered by light (photons), whereas betavoltaics are powered by beta radiation (electrons). As beta particles travel through the semiconductor, a substantial fraction of their energy is used to create electron-hole pairs (EHPs) along their path.\cite{klein1968bandgap} Significantly, since the energy of beta particles, even those produced by relatively low-energy isotopes, can be orders of magnitude higher than the semiconductor bandgap energy, each incident particle can potentially produce hundreds or thousands of EHPs. The electronic behavior of a p-i-n semiconductor junction subjected to beta radiation can be seen in Fig. \ref{band diagram}. Because the energy source for betavoltaic devices is a decaying isotope, radioactive material can be packaged with the device to produce a totally self-contained power source that can function for decades without inputs or maintenance.

The performance of a betavoltaic device depends primarily on three factors:  the identity and activity of the radioisotope source, the properties of the semiconductor material, and the geometry of the device.\cite{olsen1993review,sachenko2015efficiency} In this study, we investigate the performance of a planar GaN-based betavoltaic device designed to be powered by the isotope \textsuperscript{63}Ni.

\begin{figure}[t]
\centerline{\includegraphics[width=\columnwidth]{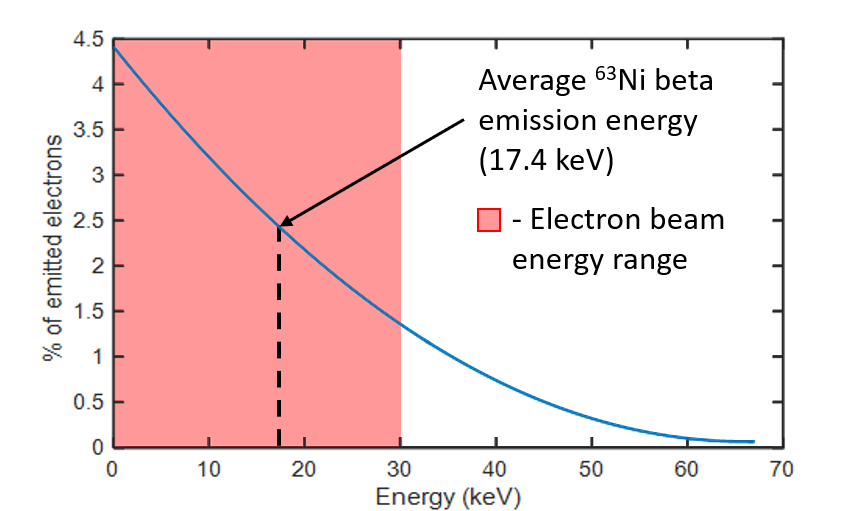}}
\caption{Beta emission spectrum of \textsuperscript{63}Ni. The curve represents the fraction of electrons emitted at each energy, the dashed line indicates the average energy of the \textsuperscript{63}Ni spectrum, and the shaded area represents the energy range covered by the electron beam used in this experiment, which had a maximum energy of 30 keV.}
\label{ni-63}
\end{figure}

\textsuperscript{63}Ni has a number of characteristics that make it ideal for use in betavoltaic devices: it has a relatively long half-life ($\sim$100 years), its beta emission spectrum is sufficiently low energy that it does not present a serious radiation safety hazard, and it does not produce alpha particles, gamma radiation or bremsstrahlung x-rays.\cite{poletiko1988determination} Moreover, as a metal, it can be incorporated directly into the architecture of a device relatively easily.  Properties of isotopes commonly considered for use in betavoltaics can be found in Table I. The \textsuperscript{63}Ni emission spectrum can be seen in Fig. \ref{ni-63}. To allow for easy variation of test parameters, in this study we used an electron beam tuned to an energy of 17.4 keV (the average energy of the \textsuperscript{63}Ni beta emission spectrum) to replicate the behavior of the isotope in place of the isotope itself.

Gallium nitride (GaN) has several advantages over Si as the semiconductor material in betavoltaic devices. The higher bandgap (3.4 eV vs. 1.12 eV) increases betavoltaic conversion efficiency, as well as making GaN significantly more robust against radiation-induced damage.\cite{son2010gan} Additionally, GaN can be alloyed with other group III elements, such as aluminum, to further increase the bandgap, and therefore the efficiency.\cite{nepal2005temperature}

\begin{table}
\caption{\label{tab:table1} Common beta-emitting radioisotopes viable for use in betavoltaic devices}
\begin{ruledtabular}
\begin{tabular}{cccc}
Isotope & Mean Energy (keV) & Max Energy (keV) & Half-Life (yrs.)\\
\hline
\textsuperscript{3}H & 5.7 & 18.6 & 12.3\\
\textsuperscript{14}C & 49 & 156 & 5710\\
\textsuperscript{35}S & 48.8 & 167.5 & 0.24\\
\textsuperscript{63}Ni & 17.4 & 66.7 & 100.1\\
\textsuperscript{90}Sr/\textsuperscript{90}Y & 195.8/933.7 & 546/2280 & 28.8/7.37e-3\\
\textsuperscript{137}Cs & 188.4 & 1176 & 30.1\\
\textsuperscript{147}Pm & 62 & 225 & 2.6\\
\end{tabular}
\end{ruledtabular}
\end{table}

The choice of GaN as a semiconductor material means that (111) Si can, with an appropriate buffer structure, be used as a growth substrate. With a procedure for growing GaN p-i-n structures on Si, it is possible to take advantage of silicon’s easy patterning to dramatically increase the surface area of the resulting betavoltaic device without increasing its wafer footprint. A wide variety of techniques exist to modify the geometry of Si substrates to increase surface area; of particular interest for this application is the wet-etching of (001) silicon to produce inverted pyramid or v-groove structures with (111)-oriented surfaces, which allow for significant surface area increases while maintaining the (111)-oriented surface needed for GaN growth.\cite{cheng2011demonstration,bower2002polymers}

\section{Radiation Modeling}

In order to maximize the efficiency of a betavoltaic device, it is important to make the depletion region large to capture as much of the energy from the penetration of beta radiation as possible, without making it so large that charge carriers are no longer able to successfully traverse the junction. 
p-i-n structures are particularly well-suited for use in betavoltaics due to the fact that their intrinsic region artificially expands the depletion region. This is significant because of the high penetration depth of even low-energy beta particles (several microns) compared to the thickness of the natural depletion region in typical GaN homojunctions (<100 nm). \cite{kozodoy2000depletion,foussekis2009photoadsorption} To verify the beta particle penetration and energy deposition depths for matching with the depletion width, modeling of radiation penetration and energy deposition in GaN was done using the Monte Carlo particle physics simulation platform CASINO.\cite{kurniawan2007investigation}

\begin{figure*}[htp]
\centerline{\includegraphics[width=2\columnwidth]{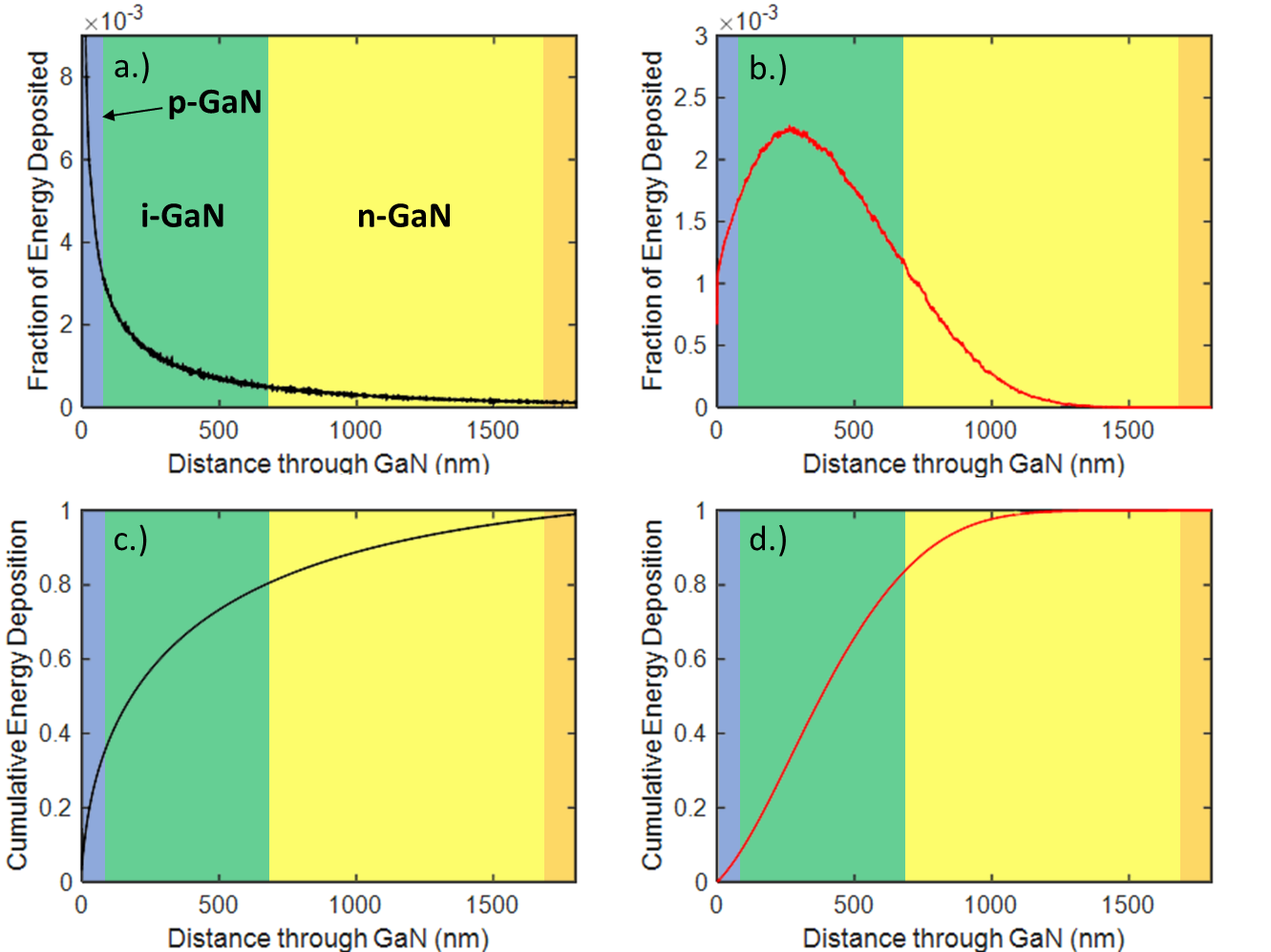}}
\caption{CASINO simulation results for beta particle penetration into GaN, with layer thicknesses shown. a.) and b.) show the energy deposition as a function of depth through GaN, whereas c.) and d.) display the cumulative fraction of the total energy deposited as a function of depth. a.) and c.) (in black) show results from a simulation using a source representing the full \textsuperscript{63}Ni beta emission spectrum, while the simulation represented in b.) and d.) (in red) used a beam source oriented perpendicular to the surface tuned to an energy of 17.4 keV.}
\label{beta sim}
\end{figure*}

Two simulation sets were run: one subjected to an isotropic particle source producing the full \textsuperscript{63}Ni beta emission spectrum to simulate the behavior of the actual \textsuperscript{63}Ni isotope, and one subjected to an electron beam source tuned to 17.4 keV (the average energy of the \textsuperscript{63}Ni spectrum) to simulate the electron beam used in this experiment. Both simulations were designed such that the incident particle count n = 50,000. These simulations show that the vast majority ($\sim$90\%) of the energy deposited by \textsuperscript{63}Ni beta radiation is deposited within or near the i-GaN layer, which means that it is deposited in the depletion region. Results of these simulations can be seen in Fig. \ref{beta sim}. In addition to the fact that these two beta sources have different energies (fixed energy at 17.4 keV vs. spectrum with average energy of 17.4 keV), they also differ in angle of incidence. The beam source is oriented perpendicular to the surface, while particles produced by the spectrum source, an isotropic emitter, can impinge upon the GaN surface at any angle between 0\textdegree and 90\textdegree. Particles entering at shallower angles do not penetrate as deeply into film, shifting the net energy deposition much closer to the material surface than would be expected from an orthogonal beam. Results of the simulated beta particle penetration were used to select the thicknesses of the layers in the fabricated device, in order to capture as much energy as possible within the depletion region.

\section{Film Growth and Device Fabrication}

\begin{figure*}[htp]
\centerline{\includegraphics[width=2\columnwidth]{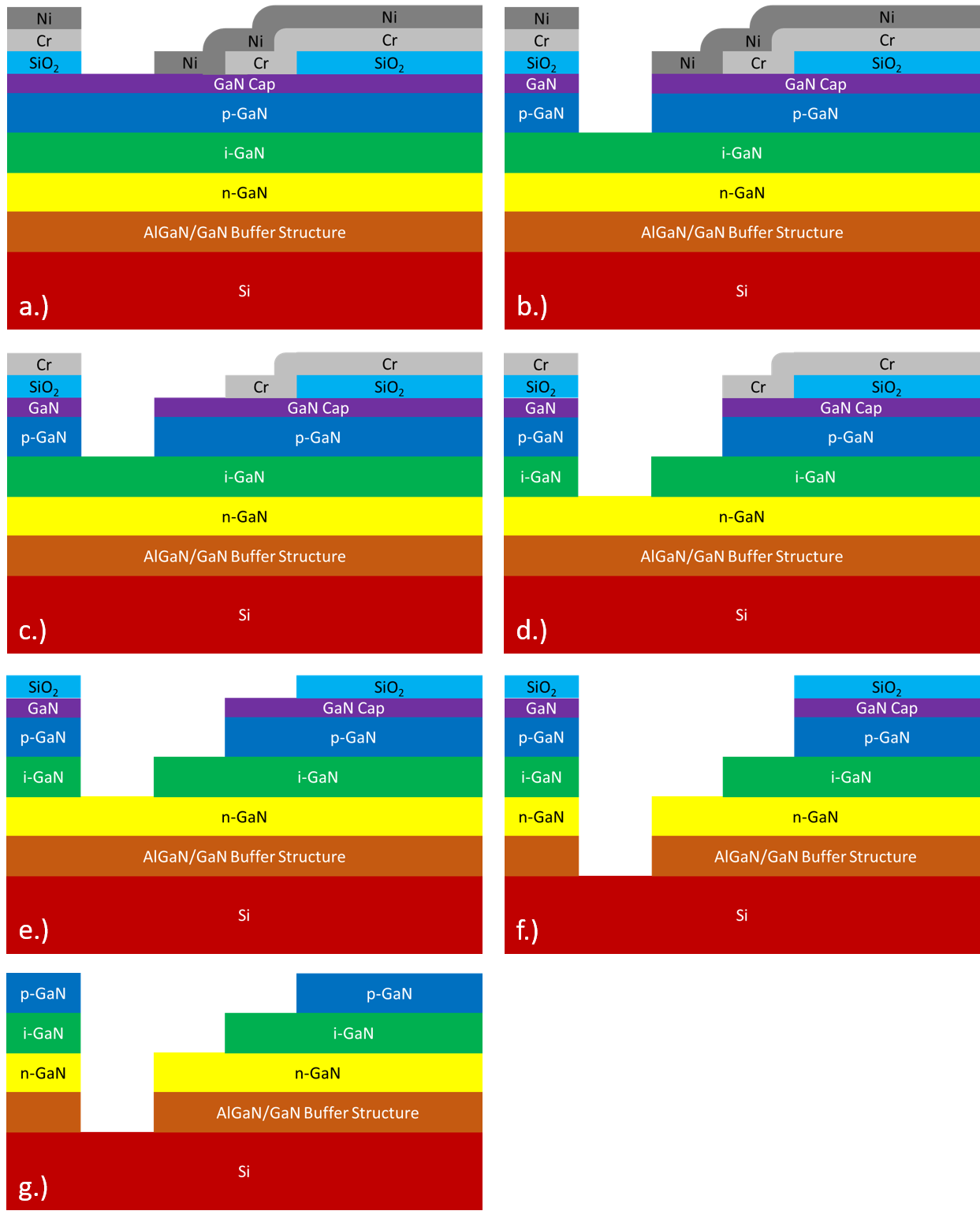}}
\caption{Triple mesa etch process steps: a.) Full AlGaN/GaN film stack with SiO\textsubscript{2}, Cr, and Ni hard masks deposited and patterned. b.) Dry etch for n-GaN patterning. c.) Wet etch to strip Ni hard mask. d.) Dry etch for i-GaN patterning. e.) Wet etch to strip Cr hard mask. f.) Dry etch for p-GaN patterning. g.) Final AlGaN/GaN architecture after removal of SiO\textsubscript{2} hard mask and GaN cap.}
\label{beta etch v2}
\end{figure*}

\begin{figure}[htp]
\centerline{\includegraphics[width=\columnwidth]{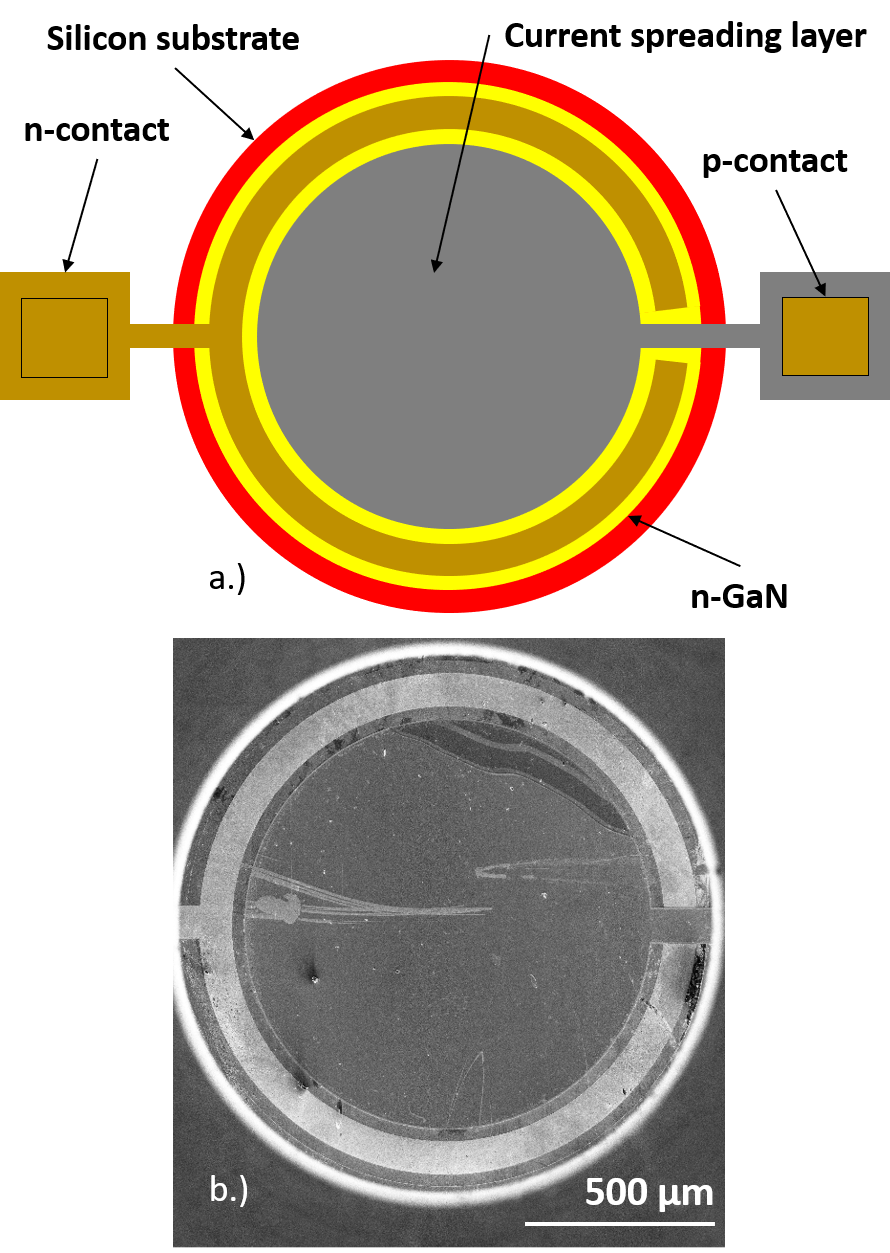}}
\caption{a.) Top-down schematic of betavoltaic device architecture and b.) Top-down SEM of fabricated betavoltaic device.}
\label{fig5}
\end{figure}

Device fabrication began with deposition of a 5-layered AlN/AlGaN/GaN buffer structure on an RCA-cleaned 4-inch (111) silicon wafer using metalorganic chemical vapor deposition (MOCVD), followed by MOCVD growth of a 3-layered GaN p-i-n structure with a 2 nm GaN cap to passivate the active layers during processing. The purpose of the buffer structure is to step smoothly from the lattice constant of (111) Si (3.84 \AA) to the lattice constant of GaN (3.19 \AA) to reduce film strain and minimize the formation of dislocations during growth. The entire stack, including the p-i-n structure, was produced in a single MOCVD process step without breaking vacuum to minimize contamination and improve lattice matching between the films. The growth details of this buffer structure have been described previously.\cite{xu2016wafer} To create the n- and p-type regions, the GaN was doped during growth with silicon and magnesium, respectively. Because GaN has a natural tendency towards unintentional n-type doping as result of the donor-like behavior, the i-region was lightly counterdoped with Mg. Based on the simulation results and comparison with existing literature, the p-GaN region had a thickness of 80 nm with a dopant concentration of 4 x 10\textsuperscript{17} cm\textsuperscript{-3}, the i-GaN region had a thickness of 600 nm with a dopant concentration of 1 x 10\textsuperscript{15} cm\textsuperscript{-3}, and the n-GaN region had a thickness of 80 nm with a dopant concentration of 3 x 10\textsuperscript{18} cm\textsuperscript{-3}.\cite{cho2003electron, zhu2012unintentional} 

The mesa etch was accomplished through a novel triple hard mask etch technique, in which three different films are deposited and patterned sequentially in order to act as hard masks for successive etches into the GaN p-i-n structure. First, a 100 nm thick SiO\textsubscript{2} film was deposited on the GaN buffer structure using plasma-enhanced chemical vapor deposition (PECVD). Then, this film was patterned to expose the areas of the p-GaN layer that were to be etched. Next, a 100 nm chromium film was evaporated onto the surface and patterned to expose the areas of the i-GaN that were to be etched. Last, a 100 nm nickel film was evaporated onto the surface and patterned to expose the areas of the n-GaN that were to be etched. 

After the three hard masks were deposited, a series of alternating dry and wet etches were done to sequentially etch layers of GaN and strip away the hard masks: First, a dry etch  was done to pattern the p-GaN, then the nickel mask was chemically removed using a metal-selective wet etch (Transene Nickel Etchant TFG, 40\textdegree C, 10 min). Second, another dry etch was done to pattern the i-GaN, then the chromium mask was removed using another metal-selective wet etch (Transene Chromium Etchant 1020AC, 40\textdegree C, 10 min.). Third, a dry etch was used to pattern the n-GaN, forming the isolation trenches around each device, after which the SiO\textsubscript{2} layer was chemically stripped in a 2\% HF solution. Finally, a very brief dry etch was done to strip the GaN cap and expose the n-GaN surface.

After the triple mesa etch, Ohmic contacts were deposited on the p- and n-type GaN layers using three separate metal evaporation and liftoff steps. The first step deposited a 8/10 nm Ni/Au (post metallization rapid thermal anneal: 470 \textdegree C, 120 s, 2:1 N\textsubscript{2}/O\textsubscript{2}, ambient pressure) current-spreading layer to facilitate the collection of charge carriers from the p-GaN layer. The second step deposited  20/100/40/80 nm Ti/Al/Pt/Au multilayer n-GaN contacts. The third step deposited the gold bond pads used to passivate the metal stacks and allow for electrical contact with the device via wirebonding. Top down schematic and SEM can be seen in Fig. \ref{fig5}. \cite{cheng2012high,khan2017design}

\begin{figure}[t]
\centerline{\includegraphics[width=\columnwidth]{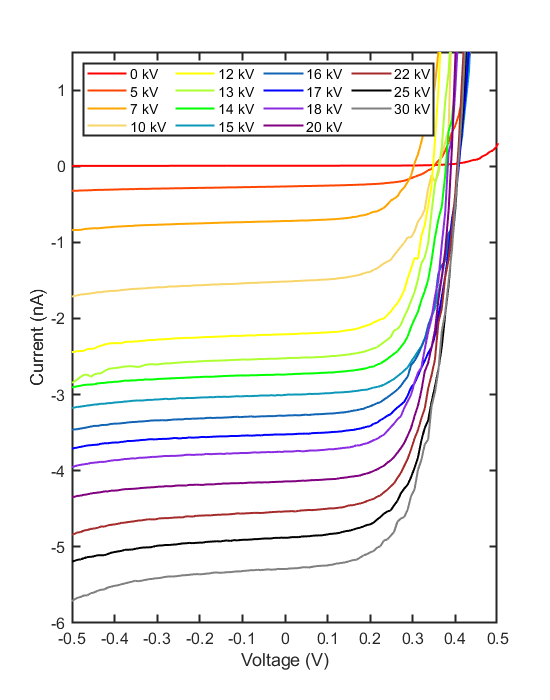}}
\caption{Betavoltaic I-V measurements under 26 nA electron beam irradiation over an accelerating voltage range of 0-30 kV.}
\label{fig6}
\end{figure}

\section{Electron beam Irradiation}

To test the devices in a radiation environment, they were subjected to electron beam irradiation from a FEI Magellan 400 XHR SEM with the accelerating voltage tuned between 0 and 30 kV and the current fixed at 26 nA. In order to conduct in-situ electrical measurements during irradiation, the devices were wirebonded to a printed circuit board (PCB) that was connected to an external measurement setup through electrical feedthrough ports built into the SEM chamber. A rectangular 995 \textmu m x  860 \textmu m window at the center of the device under test was continuously irradiated during I-V measurements, resulting in an electron flux density of 0.19 nm\textsuperscript{-2} or 3.04 \textmu A/cm\textsuperscript{2}.  The irradiation area was scanned at 88 Hz to eliminate the influence of any position-specific device defects on the measured device current. The I-V curves generated by these measurements can be found in Fig. \ref{fig6}

\begin{figure}[htp]
\centerline{\includegraphics[width=\columnwidth]{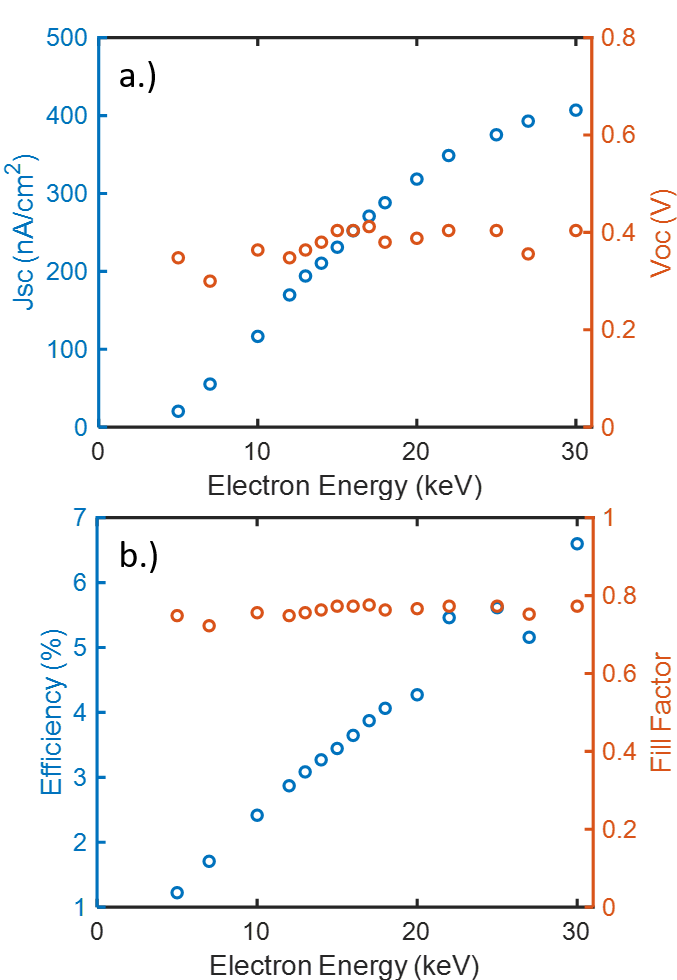}}
\caption{Betavoltaic device parameters as a function of incident electron beam energy. a.) displays short-circuit current density and open-circuit voltage, and b.) displays overall device efficiency and fill factor.}
\label{beta plots}
\end{figure}

\section{Results and Discussion}

The total device efficiencies were calculated using eqs. (1-3) and  the fill factor (FF) was calculated using eqs. (4-5), where $\eta$ is overall efficiency, $P_{out}$ is the power output of the betavoltaic device, $P_{in}$ is the power of the electron beam, $I_{m}$ and $V_{m}$ are the current and voltage at the point of maximum power, $I_{beam}$ and $V_{beam}$ are the current and voltage of the incident electron beam, and  $V_{oc}$ is the open circuit voltage of the betavoltaic device.\cite{khan2017design,sachenko2015analysis}

\begin{equation}
    \eta = \frac{P_{out}}{P_{in}}
\end{equation} 

\begin{equation}
    P_{out} = I_{m} * V_{m}
\end{equation}

\begin{equation}
    P_{in} = I_{beam} * V_{beam}
\end{equation}

\begin{equation}
    v_{OC} = \frac{{}V_{OC}}{k_{B}T}
\end{equation}

\begin{equation}
    FF = [v_{OC}-ln(v_{OC}+0.72)]/(v_{OC} + 1)
\end{equation}

Fig. \ref{beta plots} contains two plots displaying parameters of the device at different e-beam accelerating voltages. 

Analysis of the I-V measurements taken under irradiation indicate a maximum efficiency of 6.6\% and FF of 0.77. As the incident beam energy is increased, the device efficiency and the short-circuit current increase, and the $V_{oc}$ and FF, which do not depend on incident energy, remain relatively constant. The continued increases of $\eta$ and $J_{sc}$ occur because the overall amount of energy being introduced into the betavoltaic is increasing, while the incident beam energy is still within the range of the \textsuperscript{63}Ni beta emission spectrum that device was designed to effectively absorb. Although differences in device architecture, radiation source, and test methodology make direct comparison difficult, these values are comparable to those that have been reported previously for similar devices.

\begin{figure}[htp]
\centerline{\includegraphics[width=\columnwidth]{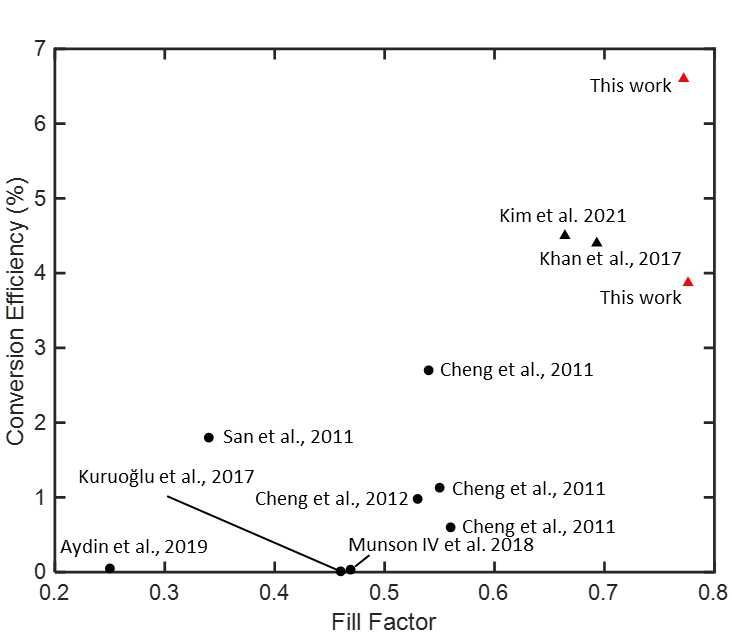}}
\caption{Comparison of results from this work with results from previously published works on GaN-based betavoltaic devices. Circles represent devices irradiated with \textsuperscript{63}Ni sources, triangles represent devices irradiated with electron beams.}
\label{beta_compare}
\end{figure}

To the best of the authors' knowledge, at the time of publication, this work represents the highest experimentally demonstrated conversion efficiency and fill factor for a gallium nitride-based betavoltaic device. A comparison of the results of this work (overall conversion efficiency and fill factor) with previous results of GaN-based betavoltaic devices from literature can be found in Fig \ref{beta_compare}.

There are a number of possible ways to optimize the device to further improve performance, in addition to growing on textured silicon substrates. One potential method includes incorporating \textsuperscript{63}Ni directly into the p-GaN contact to bring the isotope source directly into contact with the device surface, eliminating losses from electrons reflected or absorbed by the current spreading layer. Additionally, further optimizing the film thicknesses (as well as the film quality) could allow for effective capture of a higher fraction of the energy deposited by incident beta radiation.

\section{Conclusions}
In this work, we demonstrate fabrication of a p-i-n GaN-on-Si betavoltaic energy converter fabricated using a triple-hard-mask mesa etch process. Device design was informed by beta particle penetration simulations generated using Monte Carlo simulation platform CASINO.  Initial measurements indicate a maximum efficiency of 6.6\%, fill factor of 0.77, open-circuit voltage of 412 mV, and short-circuit current of 407 nA/cm\textsuperscript{2}. This conversion efficiency and fill factor are the highest experimentally demonstrated in gallium nitride to date. It is expected that the device performance could be further improved by growing the AlGaN/GaN film stack on textured Si substrates to increase the effective surface area, by optimizing the device layer thicknesses, and by integrating  \textsuperscript{63}Ni directly into the p-GaN current-spreading layer so that it is in direct contact with the device surface.

\begin{acknowledgments}

Fabrication and material characterization work were performed in part at the Stanford Nanofabrication Facility (SNF) and Stanford Nano Shared Facilities (SNSF). This material is based upon work supported by the National Science Foundation Graduate Research Fellowship [Grant No.DGE – 1147470]. M.B. would like to acknowledge the Stanford Nano Shared
Facilities (SNSF) with NSF support (ECCS-1542152) and NSF Graduate Fellowship (DGE-1656518).

\end{acknowledgments}

\nocite{*}
\bibliography{beta}
\end{document}